\documentclass{aa}  
\usepackage{graphicx}
\usepackage{natbib}
\usepackage{txfonts}
\newcommand\xmm{{\sl XMM-Newton}}
\newcommand\chandra{{\sl Chandra}}
\newcommand\nustar{{\sl NuSTAR}}

\newcommand{\lumcgs}{erg~s$^{-1}$}

\newcommand{\lognhsq}{{\rm log} [N_{\rm H}/{\rm cm^{-2}}]}

\newcommand{\nh}{cm$^{-2}$}
\newcommand{\nhi}{\rm cm^{-2}}
\newcommand{\nhsym}{N_{\rm H}}
\newcommand{\lumh}{L_{\rm 2-10}}
\newcommand{\lumhh}{L_{\rm 10-40}}
\newcommand{\msun}{M_{\odot}}
\newcommand{\lsun}{L_{\odot}}
\newcommand{\lir}{L_{\rm IR}}
\newcommand{\lbol}{L_{\rm bol}}

\def\hstwfc{{\em HST/WFC3}}
\def\wise{{\em WISE}}
\def\mic{{$\mu$m}}
\def\spitzer{{\it Spitzer}}
\def\herschel{{\it Herschel}}

\begin{document}

   \title{The hyperluminous Compton-thick  $z\sim2$ quasar nucleus of the hot~DOG W1835$+$4355 observed by \nustar}

   \titlerunning{Hyperluminous Compton-thick hot~DOG W1835$+$4355 observed by \nustar}
   
   \author{L. Zappacosta
          \inst{1}
          \and
          E. Piconcelli\inst{1}
          \and F. Duras\inst{1,2}
          \and C. Vignali\inst{3,4}
          \and R. Valiante\inst{1}
          \and S. Bianchi\inst{2}
          \and A. Bongiorno\inst{1}
          \and F. Fiore\inst{1}
          \and C. Feruglio\inst{5,6}
          \and G. Lanzuisi\inst{3,4}
          \and R. Maiolino\inst{7,8}
          \and S. Mathur\inst{9,10}
          \and G. Miniutti\inst{11}
          \and C. Ricci\inst{12,13,14}
          }

   \institute{INAF - Osservatorio Astronomico di Roma, via di Frascati 33, 00078 Monte Porzio Catone, Italy   \and
%
     Dipartimento di Matematica e Fisica, Universit\`a degli Studi Roma Tre, via della Vasca Navale 84, I-00146 Roma, Italy \and
     Dipartimento di Fisica e Astronomia (DIFA), Universit\`a di Bologna, via Gobetti 93/2, 40129, Bologna, Italy \and
     INAF--Osservatorio di Astrofisica e Scienza dello Spazio di Bologna, via Gobetti 93/3, 40129, Bologna, Italy \and
     Osservatorio Astronomico di Trieste, Via G.B. Tiepolo 11, I-34143 Trieste, Italy \and
     Scuola Normale Superiore, Piazza dei Cavalieri 7, I-56126 Pisa, Italy \and
     Cavendish Laboratory, University of Cambridge, 19 J. J. Thomson Ave., Cambridge CB3 0HE, UK \and
     Kavli Institute for Cosmology, University of Cambridge, Madingley Road, Cambridge CB3 0HA, UK \and
     Department of Astronomy, The Ohio State University, 140 West 18th Avenue, Columbus, OH 43210, USA \and
     Center for Cosmology and AstroParticle Physics, The Ohio State University, 191 West Woodruff Avenue, Columbus, OH 43210, USA \and
     Centro de Astrobiolog\'ia (CSIC-INTA), Depto. de Astrof\'isica, ESAC Campus, Camino Bajo del Castillo s/n, 28692, Villanueva de la Ca\~nada, Spain \and
     N\'ucleo de Astronom\'ia de la Facultad de Ingenier\'ia, Universidad Diego Portales, Av. Ej\'ercito Libertador 441, Santiago, Chile \and
     Chinese Academy of Sciences South America Center for Astronomy and China-Chile Joint Center for Astronomy, Camino El Observatorio 1515, Las Condes, Santiago, Chile \and
     Kavli Institute for Astronomy and Astrophysics, Peking University, Beijing 100871, China
   }
             

 
   \abstract
       {We present a 155~ks \nustar\ observation of the $z\sim2$ hot dust-obscured galaxy (hot DOG) W1835+4355. We extracted spectra from the two \nustar\ detectors and analyzed them jointly with the archival \xmm\ PN and MOS spectra. We performed a spectroscopic analysis based on both phenomenological and physically motivated models employing toroidal and spherical geometry for the obscurer. In all the modelings, the source exhibits a Compton-thick column density $\nhsym \gtrsim 10^{24}$~\nh, a 2-10~keV luminosity $L_{2-10}\approx2\times10^{45}$~\lumcgs , and a prominent soft excess ($\sim5-10$\% of the primary radiative output), which translates into a luminosity $\sim10^{44}$~\lumcgs. We modeled the spectral energy distribution from 1.6 to 850~$\mu m$ using a clumpy two-phase dusty torus model plus a modified blackbody to account for emission powered by star formation in the far-infrared. We employed several geometrical configurations consistent with those applied in the X-ray analysis. In all cases we obtained a bolometric luminosity $L_{\rm bol}\approx3-5\times10^{47}$~\lumcgs, which confirms the hyperluminous nature of this active galactic nucleus. Finally, we estimate a prodigious star formation rate of $\sim$3000~$\msun\,yr^{-1}$, which is consistent with the rates inferred for $z\approx2-4$ hyperluminous type I quasars.
         The heavily obscured nature, together with $L_{\rm bol}$, the ratio of X-ray to mid-infrared luminosity, the rest-frame optical morphology, and the host star formation rate are indicative of its evolutionary stage.
We can interpret this as a late-stage merger event in the transitional, dust-enshrouded, evolutionary phase eventually leading to an optically bright AGN.       
       }


   \maketitle
%

\section{Introduction}
Recent sensitive wide-area mid-infrared ($\sim3-30$~\mic; MIR) surveys allowed an almost obscuration-independent selection of rare populations of distant ($z=2-4$) quasars that are characterized by their huge infrared (IR) output ($\lir\gtrsim10^{14} \lsun$), which gave them the name hyperluminous infrared galaxies. 
Using the all-sky survey performed by the Wide-field Infrared Survey Explorer \citep[{\it WISE};  ][]{W2010}, samples of $\sim$100-1000 high-redshift ($z\approx2-4$) MIR-bright\footnote{Specifically at observed wavelengths $\gtrsim10$~\mic.} type I  and type II  hyperluminous sources have been selected according to specific selection criteria \citep[e.g., ][]{WED2012,E2012,wu2012}.
These rare systems are important as they may provide the clearest view of quasars at the peak epoch of AGN activity \citep[$z\approx2-3$; ][]{R2006,HRH2007,MH2008,DV2014} and in that they may be a low-redshift analog of the most luminous and massive highest-redshift quasars known \citep[e.g.,][]{F2001,F2003,will2010,B2016}. They therefore provide an important testbed to models of supermassive black hole (SMBH) formation and popular AGN/galaxy coevolution scenarios \citep{SR1998,F1999,K2003,VSH2006,H2008,CO2013,HB2014}. They may indeed represent different evolutionary phases of popular merger-driven quasar formation scenarios in which the loss of angular momentum of large cold gas reservoirs as a consequence of multiple major galaxy encounters causes rapid SMBH growth through infall of chaotic nuclear matter. This triggers powerful AGN activity and generates strong bursts of star formation \citep{S1988,DM2005,H2006,H2008,M2008,N2010,T2012}. A key transitional stage of this process predicts that the dense concentrations of infalling matter will eventually isotropically enshroud the active nucleus, heavily obscuring the sightline to the AGN and causing it to appear red and heavily absorbed at shorter than near-infrared wavelengths \citep{H2006,U2008,G2012}.

Alternative scenarios involving stochastic short transitional phases of high-matter nuclear accretion flows have also been considered for obscured quasars. They are not connected to major mergers, but are rather linked to  minor mergers and episodic cold-gas accretion episodes  \citep[flickering AGN scenario; ][]{S2012,S2015,farrah2017}.

Observationally obscured AGN are better suited for studying the AGN/galaxy coevolution because they allow the best view of the host galaxy. 
The most promising $z=2-4$ AGN candidates for the transitional dust-enshrouded phase are the so-called ``hot dust-obscured galaxies'' \citep[hot DOGs; ][]{E2012,wu2012}, that is, sources selected to be bright in the \wise\ 12 \mic\ (W3) and/or 22 \mic\ (W4) bands and faint or undetected in the 3.4 \mic\ (W1) / 4.6 \mic\ (W2)  bands \citep[hence called W1W2 drop-out; ][]{E2012}. These sources are hyperluminous with $\lbol>10^{47}$~\lumcgs, and they are rare \citep[$\sim1000$ across the sky; ][]{E2012}. They exhibit a peculiar spectral energy distribution (SED)  that peaks in the MIR, which suggests that the main source powering these objects is the central AGN and not a powerful starburst.  The temperatures derived for their dust reservoir  are indeed on the order of $T=60-100$~K \citep{wu2012,fan2016}, which is higher than the typical dust temperatures of $T=30-40$~K in other more common MIR-selected sources such as the normal submillimeter galaxies \citep[e.g., ][]{m2012} and dust-obscured galaxies \citep[DOGs; ][]{D2008}. For this reason, they have been dubbed hot DOGs.
These sources have been found in regions populated on arcminute scales ($\sim500-700$~kpc) by overdense concentrations of submillimeter galaxies, suggesting that they are indeed possible signposts of protocluster regions \citep{J2014}. 
As expected, the few X-ray observations of hot DOGs performed so far showed remarkably clearly that they are luminous and heavily obscured quasars \citep[][]{S2014,P2015,A2016,R2017,V2017}. In particular, \citet[][hereafter P15]{P2015} found from studying the \xmm\ X-ray spectrum of the hot DOG WISE~J183533.71+435549.1 \citep[z=2.298;][hereafter W1835]{wu2012} that the source is reflection dominated and hence obscured by Compton-thick\footnote{The Compton-thick obscuration is formally defined as absorption due to material with a column density of $\nhsym\geq1.5\times10^{24}$~\nh.} (CT) column densities. This provides further evidence that the source may be in the transitional heavily obscured phase.

We here report on the $\sim$150~ks \nustar\ observation of the hot DOG W1835, which provides the first and most obscured \nustar\ detection of a $z>2$ AGN. We perform a broadband $\sim$0.5-20~keV joint \xmm-\nustar\ spectral analysis and updated SED modeling.

Throughout the paper we assume a cosmology with $\Omega_\Lambda=0.73$ and $H_0=70\,\rm  km\, s^{-1} Mpc^{-1}$. Errors are quoted at $1\sigma$ and upper and lower limits at $90\%$ confidence level, unless otherwise stated. 
\section{NuSTAR data reduction}
\begin{figure}[!t]
    \label{image}
   \begin{center}
\includegraphics[width=0.5\textwidth]{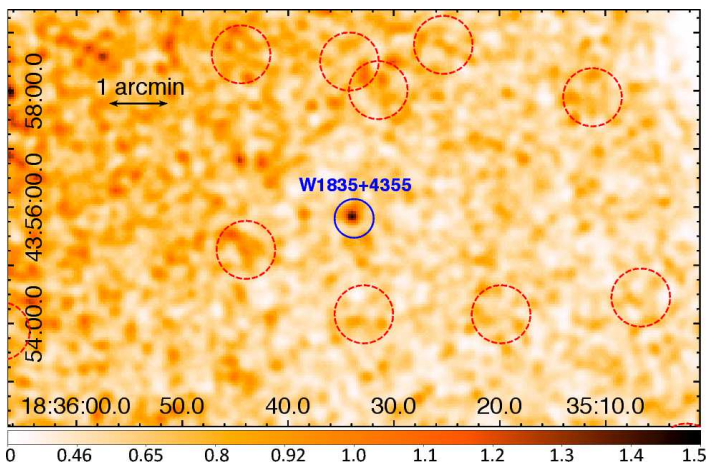}
   \end{center}
\caption{\nustar\ image in equatorial coordinates of the W1835 field in the energy range 3-24~keV (FPMA and FPMB were coadded in order to increase the signal-to-noise ratio). The blue solid 20~arcsec radius circular aperture reports the position of the hot~DOG. Red dashed circles report the regions (30~arcsec radius) centered at the position of all the \xmm- detected sources in the field (except for W1835) that were removed during the background-modeling procedure. The image was smoothed with a Gaussian kernel of 4~pixels (i.e., 10~arcsec) for better visualization. The color bar reports count/pixel values.}  
\end{figure}

The source W1835 was observed with \nustar\ \citep{H2013} for 155~ks (OBSID 60101040002) on 19 November 2015. We removed periods of high 3-20~keV background.  
The remaining cleaned event files consist of 140 and 143~ks for detectors FPMA and FPMB, respectively. Because the source is expected to be weak and \nustar\ has a spatially dependent background (at energies $<$15-20 keV) given by cosmic unfocused straylight striking the detector through the open design of the telescopes, we chose to completely model the background in the whole detector areas. We employed the {\sl nuskybdg} procedure \citep{W2014}. We sampled the background by extracting from each of the four chips in each detector spectra from circular regions of 3-4 arcmin radius. We excluded from these regions the hot DOG position, the chip gaps, and all the point sources detected in the \xmm-PN and \xmm-MOS images, adopting circular regions with an aperture of radius $\sim30$~arcsec. We then performed a joint modeling of the instrumental FPMA/FPMB and cosmic focused and unfocused (i.e., straylight) backgrounds. From the best fit we reconstructed the background image and checked visually for residual spatial variations in the background-subtracted images.
The source is detected in both images at a significance of $\sim2.5-2.7\sigma$ within 20~arcsec radius. The combined significance is $3.3\sigma$. In  Fig.~\ref{image} we report the combined 3-24~keV FPMA and FPMB images in order to highlight the significance of the hot DOG (blue circle) above the field and compare the \xmm-detected point sources in the field (dashed red circles).
We then chose the source spectral extraction radius for each detector by simultaneously maximizing the signal-to-noise ratio and the net-source counts within increasing apertures \citep[for details, see ][]{z2018} centered on the \xmm-detected source position (P15). The chosen extraction radii are 30 and 20 arcsec for FPMA and FPMB, respectively. We extracted the spectra from these circular regions using the {\sl nuproduct} task in NuSTARDAS v.~1.4.1 with calibration database (CALDB) v.~20150123. The background spectra within the source spectral extraction regions were simulated from the best-fit model with 100 times the exposure time in order to ensure good statistics for background subtraction.  

\section{X-ray spectral analysis}
The 3-24~keV FPMA and FPMB spectra consist of $40^{+7}_{-6}$ and $21^{+7}_{-5}$ background-subtracted counts (which are 24.5\% and 24.8\% of the total number of counts) whose $1\sigma$ uncertainties were estimated assuming Poissonian statistics \citep{g1986}. The apparent discrepancy between the collected counts in the two detectors is due to the different adopted spectral extraction radii as the \nustar\ PSF encloses in its 30~arcsec aperture (the  FPMA extraction region) $\sim1.5$ times more flux than in its 20~arcsec aperture (the FPMB extraction region).
We used them jointly with the XMM-PN and the coadded XMM-MOS spectra extracted and produced by  P15 from a 42~ks \xmm\ observation (OBSID 0720610101) performed on 18 August 2013. The addition of PN and MOS spectra (106 and 71 net-counts in the 0.5-8~keV band, respectively) allowed us to perform spectral modeling in the broad 0.5-24~keV band. All four spectra were grouped to a minimum of one net-count (i.e., background-subtracted) per bin.
We note that \nustar\ has $26^{+6}_{-5}$ (FPMA+FPMB) net-counts compared to $31\pm0.6$ for the XMM-PN and $27^{+6}_{-5}$ for the XMM-MOS in the common 3-8~keV energy band, and it expands the spectral coverage above the unexplored $\sim8$~keV spectral region (i.e., $\sim25$~keV rest-frame) with another 35 source counts up to $\sim$24~keV.
We used the Cash statistic (C-stat) as implemented in XSPEC v.~12.8.2 with direct background subtraction \citep[][]{C1979,WLK1979}. 

\subsection{Empirical models}\label{empirical}
In P15 the \xmm\ data alone suggested a best-fit parametrization consisting of a reflection-dominated model. The source was modeled by a cold Compton reflection component model empirically parametrized by a reflector with an infinite plane geometry and infinite optical depth, which also includes the most pronounced fluorescent lines from Fe and Ni K-shell transitions. To this model a soft component consisting of a power law with fixed photon index $\Gamma_{sc}=2.5$ was added. It accounts for the soft excess over the simple reflection model, which might originate from the primary component that leaked unaltered from the absorber or from photoionized or collisionally ionized or shocked gas \citep[e.g., ][]{GB2007,TV2010,F2013}. We refer to it as the ``scattered component''. A further Gaussian component parametrizing the presence of a unresolved ionized Fe transition at 6.7~keV was added in order to account for residuals in the high-energy part of the Fe~K$\alpha$ line at 6.4~keV. This may indicate a possible presence of an ionized reflector or a high-temperature collisionally ionized plasma surrounding the active nucleus. 
The addition of the \nustar\ data allows an energy coverage up to observed $\sim$20~keV. This enables us to cover rest-frame energies in the range $\sim$1.6-70~keV and therefore obtain better global constraints on the high-energy part of the spectrum and further shed light on the obscuration state of this quasar. 
Because of the heavily obscured nature of the source, we properly account for Compton scattering and geometric effects through our modeling.
These effects are more pronounced at the highest column densities. 

\subsubsection{Power-law-based models}\label{pow}
We first tried simple models and checked the consistency of the broadband parameterization by comparing it with the parameterization obtained by  P15 using \xmm\ data alone. In our models we always accounted for a multiplicative factor for the possible difference in the calibration between \xmm\ and \nustar\ \citep[which has been measured to be no more than 10\%;][]{MK2015} and source flux variability that may have occurred between observations. 
We started with an unabsorbed power-law model.
This resulted in a poor fit (C-stat$/dof$=174/157) with a photon index $\Gamma=0.9\pm0.1$ that is consistent with the slope obtained by modeling \xmm-only data (P15). 
The addition of a cold intrinsic absorber ({\sc zwabs} model in XSPEC) gave $\Gamma\approx1$ and $\nhsym\sim 10^{23} \nhi$ , but did not improve the modeling much (C-stat$/dof$=173/156) over the previous parameterization. Strong residuals at the position of the Fe K lines are present. These were interpreted by P15 as due to a prominent neutral Fe K$\alpha$ line at 6.4~keV with an equivalent width (EW) larger than 1-2~keV, and also due to a ionized Fe line at $\sim$6.67~keV. By fixing $\Gamma=1.9,$ we also obtained high residuals at low energies (1-2~keV). Hence we added a low-energy scattered power-law component whose photon index was tied to that of the primary component. We found a best-fit (C-stat$/dof$=169/156) with $\nhsym\approx6\times10^{23}$~\nh\ and a soft scattered component flux compared to that of the unabsorbed primary (called scattered fraction\footnote{Given the $\Gamma$ tied between the scattered and primary power-law components, we calculated it as the ratio between the model normalizations.}, hereafter $f_{\rm sc}$), which is in the $1\sigma$ range $f_{\rm sc}\approx10-20$\%. This is significantly higher than normally found for heavily absorbed AGN \citep[e.g.,][]{B2014,L2015}. Leaving $\Gamma$ free to vary did not improve our modeling signficantly. The addition of two unresolved Gaussian lines to account for the neutral and ionized Fe lines further improved our fit (C-stat$/dof$=147/154), giving a column density  $\nhsym=1.4^{+0.7}_{-0.5}\times10^{24}$~\nh                          that is compatible with CT absorption.
We further modified the absorber by accounting for Compton scattering of X-ray photons using the multiplicative {\sc cabs} model. This model accounts for scattering of photons outside of the line of sight (therefore neglecting photons that scatter into the line of sight). Compton scattering at the highest column densities is a non-negligible effect and further suppresses the level of the primary continuum at a fixed column density. This leads to an almost unaltered best-fit parameterization that we refer to as model~{\sc CAbsPow} with the exception of the scattered fraction, which resulted in a more reasonable $f_{\rm sc}=5^{+4}_{-3}$\%. The 2-10~keV  and 10-40~keV unabsorbed luminosities are $\lumh=2.9\times10^{45}$~\lumcgs and $\lumhh=1.9\times10^{45}$~\lumcgs. A summary of the parameters derived by our spectral analysis is given in Table~\ref{xrayparameters}.

\subsubsection{Modeling scattering and reflection from dense medium}
The presence of the neutral Fe line suggests the existence of an additional reflection component that still must be accounted for.  
 In addition to a primary power-law component modified by photoelectric absorption and Compton scattering, we therefore modeled the spectrum by employing  a reflection component including  spectral features from neutral Fe and Ni at 6-7~keV parameterized by the {\sc pexmon} model \citep{N2007} and a 6.65~keV (best fit from \xmm\ spectra alone, see P15) line for the ionized Fe line at $\sim$6.67~keV. The reflection model assumes an infinite planar geometry with infinite optical depth illuminated by the primary continuum and subtending for an isotropic source an angle of $\Omega=2\pi\times R$, where $R$ is the reflection parameter. In the model we assumed  solar abundance, an exponential energy cutoff for the incident primary power-law $\rm E_c=200$~keV \citep{Fa2015}, and a reflector inclination angle of 60~deg (default in {\it XSPEC}). We further added a soft-excess component parameterized as a power law with photon index tied to that of the primary component. In order to decrease the number of free parameters, we set its flux to be 2\% of the primary flux, as is commonly found in heavily obscured ($\nhsym\gtrsim10^{23} \nhi$) AGN \citep[e.g., ][]{L2015}. We obtain a best fit with a primary component completely suppressed by $\nhsym \gg 10^{25} \nhi$ and in which only the scattered and reflection components effectively contribute to the modeling\footnote{We verified that the derived column density is insensitive to the particular choice of inclination angle.}. A more accurate treatment of scattering and reflection in high dense medium is performed by exploiting physically motivated toroidal models in Section~\ref{torus}.  

\begin{figure*}[!t]
   \vspace{2cm}
   \begin{center}
\includegraphics[height=0.35\textwidth,angle=0]{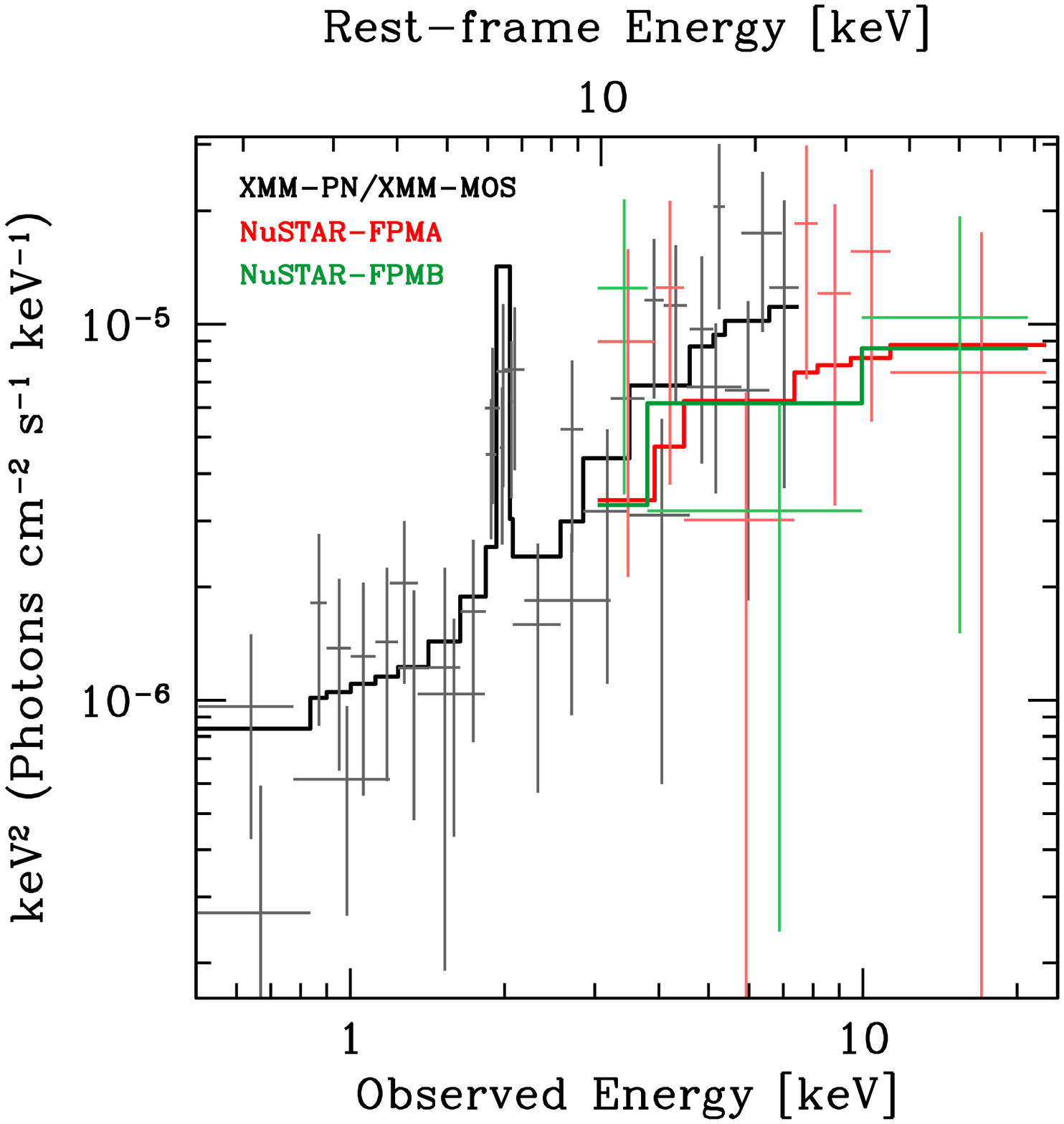}
\includegraphics[height=0.35\textwidth,angle=0]{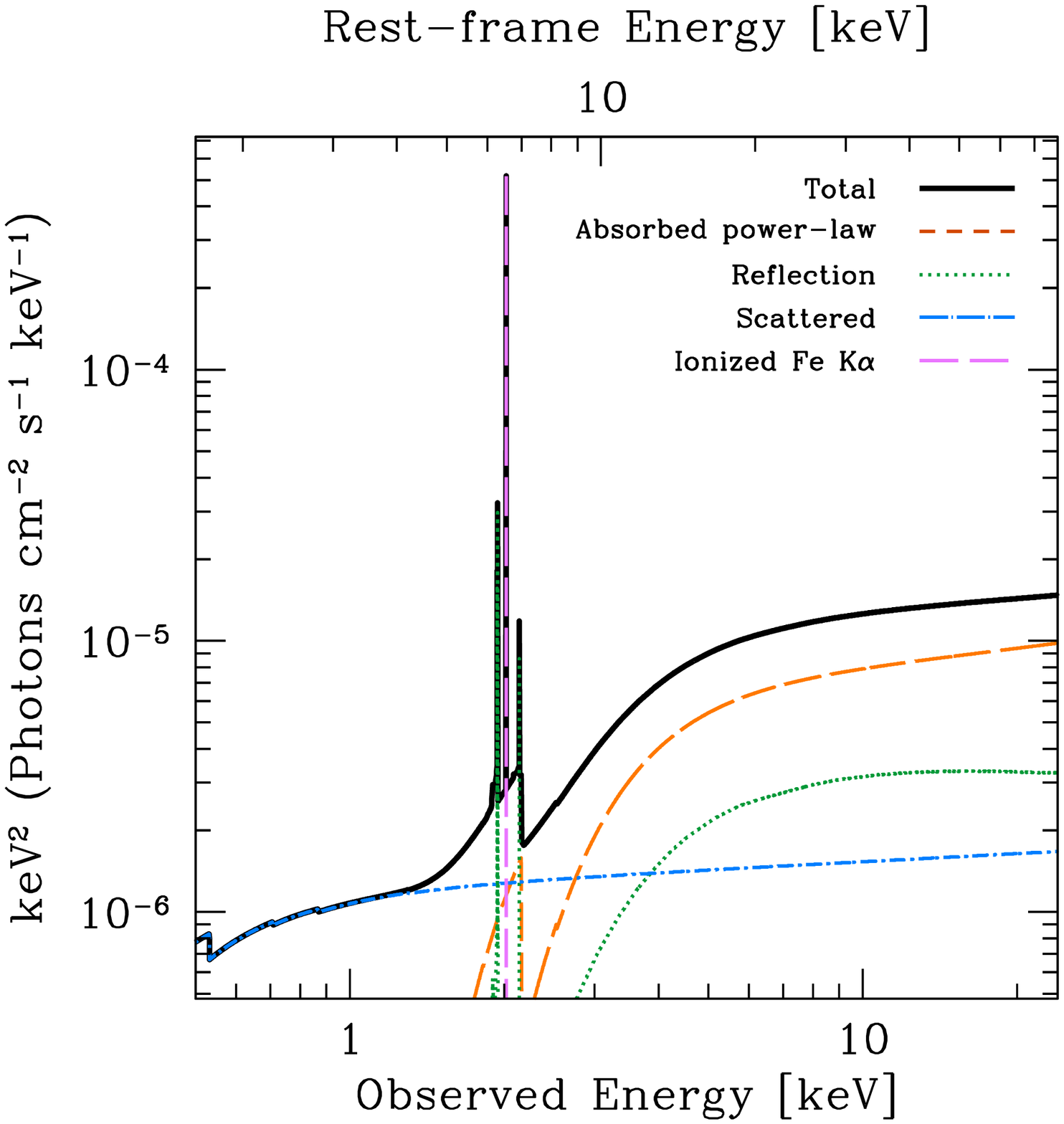}
   \end{center}
\caption{Left panel: \xmm\ and \nustar\ broadband 0.5-24~keV unfolded spectrum of W1835 grouped at a minimum of five net-counts per bin for better visual representation. The black, red, and green thick solid lines are the best-fit {\sc MYTor} unfolded model (as reported in Table~\ref{xrayparameters} and discussed in Section~\ref{torus}) for \xmm-PN, \nustar-FPMA, and \nustar-FPMB, respectively. Data are reported in lighter colors. At low energy we omit the \xmm-MOS best-fit model for clarity and report MOS and PN spectra with the same color. Right panel: Corresponding best-fit theoretical model (thick solid black line) and subcomponents (discontinuous lines). The absorbed primary power-law, the reflection, and the scattered components are reported with dashed orange, dotted green, and dash-dotted blue lines. The magenta dashed line indicates the ionized Fe line.} 
   \label{spectrum_tor}
\end{figure*}

The primary component is heavily obscured, therefore we tried to model the spectrum assuming a reflection-dominated scenario in which the coronal component is completely absorbed. Hence we removed the absorbed power-law component from our model. In our parameterization we linked the power-law slope and normalization of the {\sc pexmon} and scattered power-law. We therefore assumed that the scattered component is composed of coronal flux leaking through the obscurer unabsorbed (i.e., assuming a patchy obscurer distribution). We furthermore assumed that the scattered flux is $2\%$ of the primary, giving rise to the reflected component.  
The resulting best-fit parameterization (model~{\sc ReflDom}) constitutes an equally good yet simpler parametrization to the data with C-stat$/dof$=149/156. This confirms that the primary absorbed component is not required to parameterize our data.  Leaving $\Gamma$ free to vary does not significantly improve the parameterization  (C-stat$/dof$=148/155). The observed 2-10~keV and 10-40~keV luminosities from the reflected component are $\lumh=2.4\times10^{44}$~\lumcgs\ and $\lumhh=5.7\times10^{44}$~\lumcgs, respectively. When we assume a column density $\nhsym\approx10^{25} \nhi$ of the obscuring material, the unobscured X-ray luminosity is a factor of $\sim100$ and $\sim10$ higher, respectively.

\begin{table*}[!t]
  \begin{center}
    \caption{X-ray spectral fitting: derived parameters}
    \label{xrayparameters}
\begin{tabular}{lccccccc}
\hline
\hline
Model$^{a}$             &    C-stat$/dof$  & $\Gamma$  &   $\nhsym$                     &  $\Gamma_{sc}$           &      $f_{sc}$              &  $\lumh$              &  $\lumhh$   \\   
                       &                   &           &($10^{24}$~\nh)                 &                          &       (\%)                &   ($10^{45}$~\lumcgs)  &  ($10^{45}$~\lumcgs)  \\   
\hline
{\sc CAbsPow$^{b}$     }  &  147/154    & (1.9)     & $1.4^{+0.7}_{-0.5}$    &   ($=\Gamma$)            &   $5^{+4}_{-3} $  &   $2.9^{+3.6}_{-1.3}$                 &   $1.9^{+2.4}_{-0.9}$       \\     
{\sc ReflDom$^{c}$     }  &  149/156    & (1.9)     &    -                          &   ($=\Gamma_{pexmon}$)     &   $    (2)    $          &  $0.2^f$             &  $0.6^f$   \\    
{\sc MYTor$^{d}$       }  &  149/155    & (1.9)     & $1.1^{+0.5}_{-0.3}$    &   ($=\Gamma$)             &  $9^{+5}_{-3}$  &   $1.6^{+0.9}_{-0.5}$                 &   $1.0^{+0.6}_{-0.6}$       \\     
{\sc BNSphere$^{e}$    }  &  149/155    & (1.9)     & $0.9^{+0.3}_{-0.2}$    &   ($=\Gamma$)             &  $15^{+6}_{-4}$  &  $0.9^{+0.3}_{-0.2}$                 &   $0.6^{+0.2}_{-0.2}$       \\    
\hline
\hline
\end{tabular}
\end{center}
  \tablefoot{(a) see Sect.~\ref{empirical}, \ref{pow} and \ref{torus}\\
             (b) {\sc wabs(zpowerlw$_{sc}$+zwabs*cabs*zpowerlw+zgauss$^{neut}_{FeK\alpha}$+zgauss$^{ion}_{FeK\alpha}$)}, where {\sc zgauss$^{neut}_{FeK\alpha}$}, {\sc zgauss$^{ion}_{FeK\alpha}$} are Gaussian lines to parametrize the neutral and ionized Fe~K$\alpha$ line;\\
             (c) {\sc wabs(zpowerlw$_{sc}$+zgauss$^{ion}_{FeK\alpha}$+pexmon)};\\
             (d) {\sc wabs(zpowerlw$_{sc}$+zpowerlw*MYTZ+MYTS+MYTL)};\\
             (e) {\sc wabs(zpowerlw$_{sc}$+BNTorus)};\\
             (f) Observed luminosity not corrected for absorption.
                                                                             }
\end{table*}

\subsection{Geometry-dependent models}\label{torus}
In order to obtain more accurate and physically motivated constraints, we used {\sc MYTorus} \citep{MY2009,Y2012}, which is a Monte Carlo model based on a toroidal circumnuclear structure that absorbs and reprocesses the primary radiation self-consistently and accounts for geometric effects, Compton scattering for radiation propagating toward high column density medium, and reflection and fluorescent line emission in the reprocessed radiation. The torus has a half-opening angle of $60$~deg and is composed of uniform and neutral material. An input power-law incident primary radiation is assumed. Its implementation in {\it XSPEC} consists of three different table model components: (1) one for the attenuation of the line-of-sight radiation due to photoelectric and Compton-scattering effects ({\sc MYTZ}); (2) one to reprocess the radiation due to reflected radiation into the line of sight ({\sc MYTS}); (3) and one that calculates the contribution from scattered line emission ({\sc MYTL}). We combined all the three components by linking the column density and setting the relative normalization constants of each component to unity. We furthermore added  a power law to account for the scattered flux at soft energies and for the ionized Fe line.

 Because of the high column density and in order to comply with the geometrical requirements invoked by the standard unification schemes in which {\rm type~II} sources are seen at high inclinations, an almost edge-on view of the torus of $85$~deg was assumed \citep{B2014,L2015,z2018}.  We also fixed $\Gamma=1.9$ and tied to it the photon index for the scattered component. 
 We fixed the normalizations of  {\sc MYTS} ($A_S$) and {\sc MYTL} ($A_L$) relative to ({\sc MYTZ}) to unity and linked all the components to the same equatorial column density $N_{\rm H}^{\rm eq}$. The latter is related to the line-of-sight column density $\nhsym$ according to the inclination angle assumed and in this almost edge-on case follows the relation: $\nhsym\simeq0.98\, N_{\rm H}^{\rm eq}$.
 
\begin{figure}[!t]
   \begin{center}
     \includegraphics[height=0.45\textwidth,angle=270]{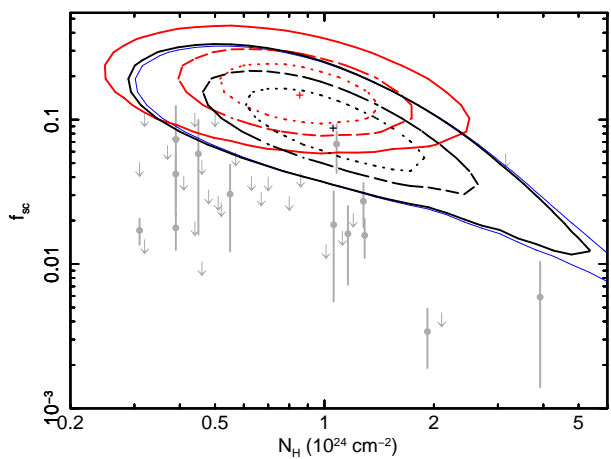}
   \end{center}
   \caption{Confidence contours for $f_{sc}$ and $\nhsym$ in {\sc MYTor} modeling (black) and the {\sc BNSPhere} model (red). Dotted, dashed, and solid lines represent contours for 68\%, 90\%, and 99\% confidence levels. Thick contours represent constraints from the joint \xmm-\nustar\ modeling. The thin solid blue contour reports 99\% constraints from \xmm-only spectral analysis. 
     Gray data indicate the scattered fractions for the heavily absorbed and CT AGN selected in the XMM-COSMOS field \citep[][errors and upper limits are 90\% c.l.]{L2015}.} 
   \label{fsnhcontours}
\end{figure}

 The best fit (C-stat$/dof$=149/155) with this model (model~{\sc MYTor}) is reported in Fig.~\ref{spectrum_tor}. In the left panel we report \xmm\ and \nustar\ unfolded spectra along with the best-fit model. The right panel shows the different subcomponents contributing to the incident model. The column density estimated with this model is $\nhsym=1.1^{+0.5}_{-0.3}\times10^{24}~\nhi$. The scattered fraction is measured to be $f_{sc}=9^{+5}_{-3}$\%, which, although slightly on the high side, is consistent within the errors with the typical fractions for highly absorbed ($\nhsym>10^{23}~\nhi$) AGN. This fraction in our parameterization is highly degenerate with the column density. This is shown by the confidence contours reported in Fig.~\ref{fsnhcontours} for the two interesting parameters\footnote{We note that the assumed high-inclination angle is conservative in terms of $\nhsym$ as it gives lower values than lower inclination angles. Furthermore, it does not affect the maximum allowed value of the scattered fractions, but it affects its constraints at the lower-end values. For instance, assuming an inclination angle of 70~deg, we obtain a 90\%  lower limit of $\nhsym\sim7\times10^{23}$~\nh\ and $f_s=0.7-22$\% ($1\sigma$ range). However, as for all inclinations angles $\lesssim$70-80~deg, the derived uncertainties are affected by the maximum tabulated $\nhsym=10^{25}$~\nh\ in the {\sc MYTorus} model, which prevents an accurate and correct sampling of the entire parameter space of interest.}. 
 Given this parameterization, the 2-10~keV and 10-40~keV unabsorbed luminosities are $\lumh=1.6\times10^{45}$~\lumcgs\ and $\lumhh=1.0\times10^{45}$~\lumcgs, respectively.
 
This model was not previously evaluated in P15 for the \xmm\ data alone. The inclusion of the \nustar\ data allows us to confirm the \xmm\ estimated values (which crucially depend on the level of the \xmm\ higher energy bins, which, being in a regime of low signal-to-noise ratio, may suffer from systematics) and to better define the upper-end value on $\nhsym$ as reported in Fig.~\ref{fsnhcontours}.

Hot DOGs are usually considered a transitional dust-enshrouded/highly star-forming phase in the merger-driven quasar formation scenario \citep{S1988,H2008,wu2012}. In this scenario, the accretion of matter proceeds through intense and chaotic accretion phases caused by the loss of angular momentum resulting from major mergers as opposed to more moderate accretion states typical of the secularly evolving planar geometries (disk-torus structure) invoked in unified models \citep[e.g.,][]{up1995}. In this case, we expect that in the former the  obscuring material is distributed more isotropically than in the toroidal structures that are typically invoked for more standard, Seyfert-like sources. If this is the true scenario for W1835, it is reasonable to evaluate a model in which the obscurer covers the entire sphere, regardless of its distance from the nucleus.  We employed a specific table model derived from Monte Carlo based calculations for a homogeneous toroidal obscurer modeled as a sphere in which there is a biconical cavity with variable opening angle \citep[hereafter {\sc BNTorus}; ][]{BN2011}. We used the case with an opening angle of $0$~deg (i.e., no biconical openings) and hence with a spherical obscurer distribution (i.e., isotropic obscuration; model~{\sc BNSphere}). This model, as well as {\sc MYTorus}, assumes a homogeneous matter distribution, but this is a probably good approximation of the spherical high covering factor obscuration scenario, even if there may likely be inhomogeneities in the gas/dust distribution (possibly with some sightlines exhibiting Compton-thin column densities). 
This model is simpler than {\sc MYTor} as it involves the same number of free parameters and does not need any inclination angle assumption. 
We obtain a best-fit (C-stat/dof=149/155) column density of $\nhsym=0.9^{+0.3}_{-0.2}\times10^{24} \nhi$ with a slightly higher scattered fraction $f_{sc}=15^{+6}_{-4}$\% (assuming $\Gamma=1.9$ and $Z=1\,Z_\odot$). Confidence contours for this case are reported in red in Fig.~\ref{fsnhcontours}. The degeneracy of the two parameters is less pronounced than for the {\sc MYTor} case, but the scattered fraction is rather high, as shown by the comparison with  highly obscured/CT AGN candidates reported from the COSMOS survey in \citet{L2015}. However, the increase in $f_{sc}$ is expected because in {\sc BNSphere} the low-energy nuclear radiation can not escape or scatter outside of the nuclear region without being obscured. Therefore, if the source truly is dust enshrouded, the soft-excess emission must come from elsewhere. The derived parameters for both {\sc MYTor} and {\sc BNSphere} modelings are reported in Table~\ref{xrayparameters}.

We expect that most of the CT absorption occurs in the innermost galactic regions around the AGN,
as argued by \citet{bb2017}. However, as stressed by these authors, their study is valid for the AGN population at large, and peculiar/rare
sources such as hot DOGs are therefore not sampled. This means that we cannot make any strong conclusion on the location of the CT absorber in W1835.

\begin{table}[!t]
  \begin{center}
    \caption{Photometric points of W1835}
    \label{photometry}
\begin{tabular}{lc@{ \mic}cc@{$\,\pm\,$}cc}
\hline
Instrument & \multicolumn{2}{c}{Band} & \multicolumn{2}{c}{Flux Density} \\
\hline
\hline                      
\hstwfc~F160W    & 1.6  & &   12.8  & 0.9    $\mu$Jy  \\
\spitzer\ IRAC1  & 3.6  & & 51.5  & 2.2  $\mu$Jy      \\
\spitzer\ IRAC2  & 4.5  & & 142.8 & 3.0   $\mu$Jy     \\
\wise\ W3        &  12  & & 6790  & 190   $\mu$Jy     \\ 
\wise\ W4        &  22  & & 24.6  & 1.0   mJy         \\
\herschel\ PACS  &  70  & & 53.6  & 4.2   mJy          \\
\herschel\ PACS  & 160  & & 92.8  & 17.1   mJy         \\
\herschel\ SPIRE & 250  & & 81.0  & 8.1   mJy          \\
\herschel\ SPIRE & 350  & & 72.0  & 7.2   mJy          \\
\herschel\ SPIRE & 500  & & 33.0  & 3.3   mJy          \\
{\em SCUBA-2}    & 850  & &  8.0  & 1.5   mJy         \\ 
\hline
\hline
\end{tabular}
\end{center}
\tablefoot{References: All photometric points are from P15, except for {the \textit{HST}/WFC3} point \citep{farrah2017} and the \herschel\ PACS points (IRSA)}
\end{table}

\section{Modeling the optical-to-infrared SED}\label{SEDW1835}
In order to have the most complete view of W1835, P15 performed SED fitting based on ten photometric points covering the MIR to far-infrared (FIR) wavelength range collected from the literature. Their SED modeling included both a galactic starburst and nuclear emission components. In particular for the latter, they used a grid of smooth torus models with a flared-disk geometry \citep{F2006,F2012}. The galactic component at longer wavelengths was well parameterized with either the $\rm{Arp}\,220$ template or with a optically thin approximation of a modified blackbody with fixed emissivity index $\beta=2$. From the latter, the authors obtained a temperature $T_{dust}\approx40~K$ for the dust component. This temperature is lower than those estimated by \citet{fan2016} and \citet{wu2012}, that is, $T_{dust}\approx70~K$ and $T_{dust}\approx90~K$, respectively. Furthermore, it is lower than the typical hot DOG temperatures and compatible with the highest temperature for normal DOGs and with the maximum temperatures derived by dust heated by stellar processes \citep[e.g.,][]{m2012,wu2012}. 
Here we perform an improved SED fitting using updated photometric points from the {\it Herschel} catalog in the NASA/IPAC Infrared Science Archive (IRSA) and adding \hstwfc\ F160W band photometry at $1.6$~\mic\ with refined modeling that accounts for the clumpy circum-nuclear torus. Table~\ref{photometry} reports the photometric points considered for the SED modeling.
\begin{table*}
  \begin{center}
    \caption{SED modelings: derived quantities}
    \label{sedparameters}
\begin{tabular}{lccccccccc}
\hline
Model                     & $\chi^2_r$ & $\theta_{oa}$     & $i$      &  $T_{dust}^a$  & $\lir^b$                  & $\lbol$              & $L_{6\mu m}$              &     SFR                 & Dust mass$^c$\\
                          &            & (deg)            & (deg)    &  (K)        & (\lumcgs)                 & (\lumcgs)            & (\lumcgs)                &     ($M_\odot\, yr^{-1}$)  & ($10^8 M_\odot$)\\
\hline
\hline
Best fit                  &  1.44      &   30             &  70      &   69    & $7.3\times10^{46}$             & $4.39\times10^{47}$    &   $1.13\times10^{47}$    &   $3300^{+100}_{-100}$     &  3.9   \\
{\sc MYTorus}-like        &  2.06      &   (60)           &  70      &   63    & $6.0\times10^{46}$             & $4.71\times10^{47}$    &   $1.07\times10^{47}$    &   $2700^{+200}_{-150}$     &  3.2   \\
{\it Sphere80}            &  4.51      &   (10)           &  (80)    &   69    & $7.3\times10^{46}$             & $5.32\times10^{47}$    &   $1.42\times10^{47}$    &   $3300^{+200}_{-200}$     &  3.9   \\
{\it Sphere90} (edge-on)  &  6.56      &   (10)           &  (90)    &   67    & $6.9\times10^{46}$             & $5.13\times10^{47}$    &   $1.38\times10^{47}$    &   $3100^{+100}_{-100}$     &  3.7  \\
\hline
\hline
\end{tabular}
\end{center}
  \tablefoot{Quantities reported in parentheses have been assumed fixed during the modeling.\\
  (a) The typical uncertainty on the temperature is $\pm5$~K.
    (b) Relative to the dust emission component.
  (c) Calculated as in \citet{B2006}.}
\end{table*}
The approach we followed is based on a two-component fitting procedure recently developed and employed in type I hyperluminous sources by \citet{D2017}. In this procedure the SED is fit with a combination of quasar and host galaxy templates.
The AGN component is described as the superposition of the accretion disk  emission and of the radiation coming from the dusty torus. We built a library of templates for the quasar emission using the broken power-law description by \citet{feltre2012} for the accretion disk and the model by \citet{stalevski2016} for a clumpy two-phase dusty torus characterized by high-density clumps embedded in a low-density and smooth medium. The two were appropriately normalized to preserve the energy balance between the ultraviolet and the IR bands. The library is composed of about 7200 templates with different values of the optical depth at 9.7~\mic, the inclination along the line of sight (from 50 to 90 degrees, in order to force a type II configuration in which the direct view of the SMBH is blocked by the torus) and the dust geometry and distribution. 
The second component of the model is a modified blackbody that accounts for the emission powered by star formation that is absorbed and then re-emitted by the surrounding dust in the MIR and FIR bands.
We parameterized it as
$$
S_{\lambda}\propto(1-e^{-\tau_\lambda}) B_\lambda(T_{dust}),
$$
where $B_\lambda(T_{dust})$ is the blackbody model and $\tau_\lambda=(\lambda_0/\lambda)^\beta$, is the optical depth, which is assumed to have a power-law dependence, with $\lambda_0$ being the wavelength at which the optical depth reaches unity and $\beta$ being the dust emissivity index. 
Because we only have a small number of photometric points, we fixed $\beta=1.6$, which seems to be the most reliable value for both local and high-z quasars \citep[][]{B2006,B2017} and $\lambda_0=125$~$\AA$ \citep{H2014,fan2016}. The cold-dust library consists of 70 templates with a range of temperatures from 30 to 100 K.

\begin{figure}[!t]
    \label{sed}
   \begin{center}
\includegraphics[width=0.5\textwidth]{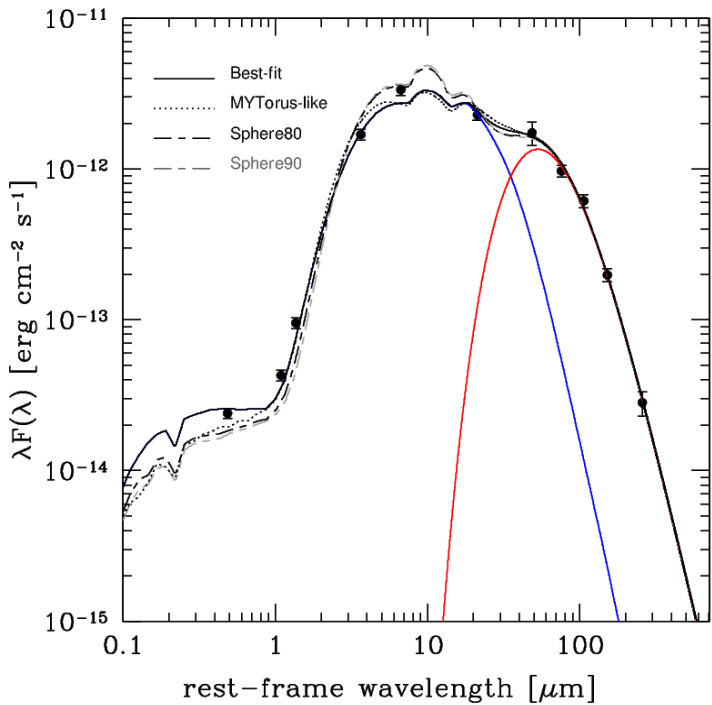}
   \end{center}
\caption{SED modeling for W1835. The solid line reports the best-fit model (black, red, and blue represent the total, AGN-only, and modified blackbody components). Dotted and dashed lines represent best-fit models assuming a toroidal geometry similar to the {\sc MYTor} and {\sc BNSphere} (black and gray show $80$~deg and $90$~deg inclination angles) models used for the X-ray analysis (Sect.~\ref{torus}). See Sect.~\ref{SEDW1835} and Table~\ref{sedparameters} for further details.} 
\end{figure}

In Fig.~\ref{sed} we show the rest-frame optical-to-FIR SED with the best-fit model (solid line).
The accretion disk plus torus emission is shown in blue and the cold-dust emission in red. The black line is the total best-fit model emission.
From this best fit we were able to derive some physical quantities of both the host galaxy and the nuclear source. They are reported in Table~\ref{sedparameters}. The intrinsic quasar bolometric luminosity was computed by integrating the emission coming from the AGN component from 1 up to 1000~\mic\ and rescaling it by a factor $\sim1.7$ to account for geometry and anisotropy of the radiation field caused by the obscurer geometry and its orientation \citep{P2007}. This correction is an average value computed for a sample of quasars characterized by a toroidal obscurer and moderate degree of obscuration ($\lognhsq\approx10^{22}-10^{23.4}$~\nh). If the emitting dust is also responsible for the CT obscuration and if a spherical geometry is in place, we should expect larger corrections and therefore possibly higher bolometric luminosities.
The IR luminosity ($L_{IR}$) of the host, a tracer of the reprocessed UV stellar light, is obtained by integrating the  dust emission component in the range 8-1000~\mic\ and is $L_{IR}=7.3\times10^{46}$~\lumcgs. From this we computed the star formation rate (SFR) using the relation by \citet{K1998} scaled to a Salpeter initial mass function, gaining an extremely high value of SFR of about $SFR=3300\pm100$~$M_{\odot}\, yr^{-1}$. The best-fit dust emission shows a temperature of about $T_{dust}=69$~K, in good agreement with the estimates by \citet{fan2016}. Our estimates differ substantially from the $SFR\approx2100$~$M_{\odot}\, yr^{-1}$ and  $T_{dust}\approx40$~K reported by P15. We verified that this difference is mainly driven by the adoption of the modified blackbody model that in P15 was approximated assuming the optically thin regime and adopting $\beta=2$. The inferred dust mass of $(3-4)\times10^8 M_\odot$ \citep[estimated as in][see Table~\ref{sedparameters}]{B2006} is consistent with typical abundances in other hot DOGs \citep{fan2016} and in coeval and higher redshift type~I analogs \citep[e.g.,][]{B2006,Valiante2014,D2017}.

In order to be consistent with the toroidal models used in our X-ray analysis, we tried to model the SED by forcing similar torus parameters. To approximate the MYTORUS geometry (i.e., {\it MYTorus-like}), we fixed the same torus opening angle and allowed inclinations in the range 70-90~deg. We also tried an almost fully covered obscurer (i.e., {\it Sphere80}) by adopting the templates with only a polar cap of 10~degrees left open and a 80-degree inclination (almost edge-on)\footnote{Compared to the smooth toroidal models used in our X-ray analysis, those used in the SED fitting are clumpy. Furthermore, the disk geometry for the latter is a conical torus (flared disk), while in MYTorus, a donut-shaped torus is assumed.}. We obtained reasonable parameterizations although with a slightly worse $\chi^2$.
The very high obscuration of the quasar and the {\it HST/WFC3 F160W} image \citep{F2016b,farrah2017}, which shows a somewhat extended and irregular host emission, means that the emission shortward of $\sim1$~\mic\ is very likely partially contributed by the host. In our models we tried to account for a maximum contribution from the AGN that does not include the host. 
Temperatures and derived SFR are in the range $T_{dust}\approx60-70$~K and around $SFR\approx3000$~$M_{\odot}\, yr^{-1}$, respectively, hence not very dissimilar from the estimate from the best-fit parameterization (see Table~\ref{sedparameters} for detailed estimates). In Fig.~\ref{sed} we report the best-fit parameterizations for the {\it MYTorus-like} (dotted) and {\it Sphere80} (dashed) cases. 
We tried a fully covered (4$\pi$) obscurer (i.e., {\it Sphere90}) and report it as the gray dashed line in Fig.~\ref{sed}. The spherical parameterizations slightly underestimate the photometric points around of $1$~\mic\ rest-frame. This is expected, and because of the extremely obscured nature of this source, these points can therefore be interpreted to be contributed by the stellar emission from the host, which is not accounted for in our modeling.

In the {\it HST/WFC3 F160W} image a low surface brightness contribution from the stellar host is visible \citep{F2016b,farrah2017}. To infer its level, we included an additional component using $\sim900$ galaxy templates from \citet{BC2003}, with different levels of extinction spanning the range $\Delta E(B-V)=0-0.5$ and assuming a \citet{CH2003} initial mass function. We find a fractional contribution of $f_{host}\approx58-75$\% (90\% confidence level interval) in the optical rest-frame band (i.e., in the Johnson B band). 
  Despite this, the global AGN and dust properties remain remarkably consistent with the values reported in Table~3. However, the {\it HST/WFC3 F160W} photometric point is the only optical constraint to the host template in the SED modeling. 
  At shorter wavelengths, the uncertainties at 90\%  level in $f_{host}$ become much larger. In addition, systematics may likely be affecting the derived value. More data at bluer rest-frame wavelengths are required to remove possible systematics and reduce the uncertainty on the host contribution.

\section{Discussion}
\subsection{Confirming high obscuration and high luminosity}\label{confirmlargeNh}
Our analysis confirms that W1835 hosts a luminous and heavily obscured quasar. 
An empirical parameterization accounting separately for different primary and reflection components suggests mild to heavy CT obscuration. A model with a reflection-only component
can already provide an excellent description of the data. Physically motivated models implementing a toroidal or spherical geometry for the obscurer and accounting for Compton-scattering effects give column densities at around $\sim10^{24}$~\nh, that is, somewhat lower than the formal threshold defining a source CT (i.e., $\nhsym=1.5\times10^{24}$~\nh). However, (i) they require scattered fractions of $f_{sc}\gtrsim5-9$\%, which is slightly on the high side but still consistent within the errors, except for {\sc BNSphere} ($f_{sc}\approx15$\%), with the canonical observed few-percent values (see Sect.~\ref{softexcess} for further details) and (ii) the scattered fractions and column density are anticorrelated. A CT obscuration with more moderate fractions is therefore very likely (i.e., at $1\sigma$ level, see Fig.~\ref{fsnhcontours}).
To our knowledge, W1835 is the most obscured, most luminous high-redshift AGN detected by \nustar. 
Other known hot DOGs whose low-energy X-ray spectra were analyzed exhibit values of $\nhsym\approx6-7\times10^{23}$~\nh\ , although with large uncertainties \citep{A2016,R2017}. An exception is W0116–0505 (hereafter W0116), one of the X-ray brightest hot~DOGs that has a column density similar to that of W1835 \citep{V2017}. Recently, \citet{G2018} analyzed the X-ray emission of a sample of extremely red hyperluminous quasars (ERQ) at $z=1.5-3.2$, finding indication of heavy absorption ($\nhsym\gtrsim10^{23}$~\nh).  Through spectral analysis of the stacked spectrum of the most weakly detected \chandra\ sources, they estimated column densities on the order of $\nhsym\sim10^{24}$~\nh. 

The derived unabsorbed luminosities for W1835 are on the order of $L_X\approx10^{45}$~\lumcgs\ (see Table~\ref{xrayparameters}). Based on this and on the estimated bolometric luminosities (Table~\ref{sedparameters}), we can derive the  bolometric corrections $k_{bol,X}=L_X/L_{bol}$ in the 2-10~keV and 10-40~keV bands. By assuming as fiducial models the {\sc MYTor} X-ray parameterization and the best-fit SED modeling, we obtain $k_{bol,2-10}\approx270$ and $k_{bol,10-40}\approx440$. If we consistently (to the X-ray model) assume a {\sc MYTorus}-like toroidal structure for the SED modeling, we obtain similar values, that is, a factor $\sim1.1$ larger. These values are all in agreement with bolometric corrections found for X-WISSH sample of $z=2-4$ type I hyperluminous MIR/optically selected quasars \citep[e.g.,][]{M2017} and with the extrapolated trend from less luminous type~II AGN \citep[e.g.][]{L2012}.

Luminous optically bright quasars have been found to exhibit X-ray luminosities weaker than those inferred by a linear extrapolation from X-ray selected lower-luminosity AGN \citep{G2009,M2015} and are more in line, but still weaker, than extrapolations from  dust-obscured galaxies at lower luminosities \citep{F2009,L2009}. The X-WISSH quasars, which are the most luminous type I AGN in the universe, clearly show this X-ray weakness compared to their MIR radiative output \citep{M2017}.
  Recently revised nonlinear relations between X-ray and MIR intrinsic luminosities have been derived by \citet{SD2015} and \citet{C2017} in order to account for the observed X-ray weakness exhibited at the highest luminosities.
\citet{R2017}, considering a sample of X-ray detected hot DOGs for which spectroscopic analysis is possible (including W1835), reported $L_{2-10}$ significantly lower than those reported by the X-WISSH sources that exhibit comparable $L_{6\mu m}$. This has been considered as evidence of further intrinsic X-ray weakness or significant underestimation of the column density. A somewhat opposite behavior has been reported by \citet{V2017} for the hot DOG W0116, for which $L_{2-10}$ appears to be comparatively higher than the extrapolated trends from hyperluminous quasars (i.e., somewhat X-ray louder). As reported by \citet{V2017}, $L_{6\mu m}$  for this source might be underestimated given that it has been computed under the assumption of isotropic emission \citep{Tsai2015,V2017}. In order to understand the level of underestimation and to consistently compare W0116 to our results for W1835, we performed SED fitting for W0116 using the photometric points provided by \citet{Tsai2015} (see their Table~2). From our best-fit model we obtain $L_{6\mu m}=1.5\times10^{47}$~\lumcgs, that is, a factor $>4$ higher than the values used by \citet{V2017}. With this new estimate we find W0116 in line with the X-ray-to-MIR luminosity values reported for type I quasars. Similarly, the spectral analysis on the X-ray stacked spectrum of weak ERQs performed by \citet{G2018} gives a value of the intrinsic X-ray luminosity that agrees with the expected $L_{6\mu m}$ for hyperluminous type I AGN. Even in this case, anysotropic MIR emission could cause an underestimation of the true MIR luminosity. In this case, we would expect these sources to exhibit slightly lower X-ray radiative outputs (by a factor of a few) than the higher MIR luminosities, as suggested by \citet{G2018} themselves.

\begin{figure}[!t]
     \vspace{2cm}
   \begin{center}
     \includegraphics[height=0.35\textwidth,angle=0]{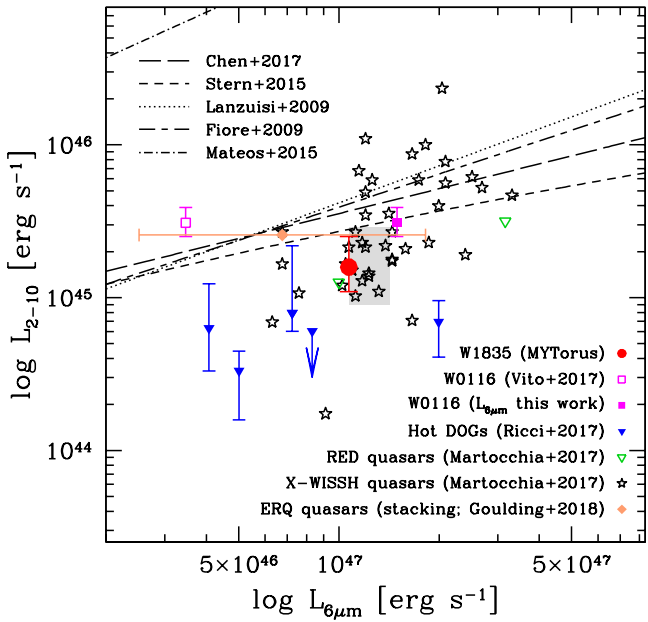}
   \end{center}
   \caption{$L_{6\mu m}$ vs. $L_{2-10}$ relation. The red circle represents W1835 for the MYTorus geometry (both in X-ray and SED analysis) with the range of systematic uncertainty due to different modelings (see Tables~\ref{xrayparameters} and~\ref{sedparameters}) reported as the gray shaded region. Blue triangles report a compilation of hot DOGs in \citet{R2017} except for W1835, which we update here. Magenta squares report W0116+0505 with $L_{6\mu m}$ reported by \citep{V2017} (empty square) or estimated by our SED fitting (filled square; see Section~\ref{confirmlargeNh}).  Black stars and green triangles represent the X-WISSH hyperluminous type I quasar sample \citep{M2017} and two reddened quasars \citep{M2017}. The orange diamond reports the best-fit $L_{2-10}$ value obtained from X-ray spectral analysis performed on a stacked X-ray spectrum of a sample of ERQ sources observed in \chandra\ in \citet{G2018}. In the latter, the error bar represents the range of $L_{6\mu m}$ in the sample.
     We also report X-ray-to-MIR relations derived for different optical/MIR/X-ray selected AGN samples.} 
   \label{lxlmir}
\end{figure}

In Fig.~\ref{lxlmir} we report the X-ray-to-MIR luminosity for the W1835 (red circle), the \citeauthor{R2017} hot DOGs sample (blue filled triangles), W0116 (magenta squares empty and filled, the latter being plotted with our SED-derived $L_{6\mu m}$), the X-WISSH sources (stars), two reddened quasars \citep[green empty triangles; ][]{F2014,B2015,M2017}, and the measure from the stacked ERQ spectrum by \citet[][]{G2018} (orange diamond). We also report X-ray-to-MIR relations obtained from MIR-selected AGN \citep{L2009,F2009}, X-ray selected AGN \citep{M2015,C2017}, and specifically tuned for the hyperluminous regime \citep{SD2015}. We note that W1835 lies higher than the average hot DOGs locus found by \citet{R2017} and  close to W0116 and the heavily absorbed ERQs and is consistent with the lower portion of the type I quasars and the red quasars. Hence it exhibits a lower degree of X-ray weakness, at least similar to that inferred for type I luminous AGN. 

\subsection{Origin of the soft excess}\label{softexcess}
In low-count spectra, the soft-excess component is usually reproduced by a single power-law component. If forced to have the same photon index as the primary intrinsic continuum, this component parameterizes an absorber in which the primary flux leaks (or Thomson scatters) unaltered through it. This parameterization is a convenient way to include a soft component by adding only one free parameter to the model and have a reference measure of $f_{sc}$. 
Given the highly obscured nature of W1835, the relatively high soft-excess emission may include additional components that are not directly linked to the quasar emission itself.

 From our SED fitting we find that the host of this AGN harbors powerful stellar nurseries capable of producing stars at a rate of $\sim$3000~$M_\odot yr^{-1}$. These levels of star formation are typical of other hyperluminous high-redshift quasars \citep[e.g.,][]{fan2016,D2017}. 
 The scattered flux may therefore be contributed by X-rays from powerful star-forming regions such as are found in local galaxies \citep{R2003}. If we assume $SFR=2000-3000$~$M_\odot\, yr^{-1}$ , we can estimate the expected 2-10~keV luminosity using the \citet{R2003} relation \citep[updated by][]{KE2012} and obtain $\lumh^{SF}\approx1.2\times10^{43}$~\lumcgs\ , which is $\sim$10\% of the estimated scattered flux (Table~\ref{xrayparameters}). This does not significantly account for the  soft excess, especially for the {\sc BNSphere} model, whose estimated column density depends very little on $f_{sc}$ (see Fig.~\ref{fsnhcontours}). Therefore the {\sc BNSphere} model is not an adequate parameterization for the obscurer. This would imply either a patchy $4\pi$ obscurer with different properties from the more local sources (in terms of geometry, covering factor, and line-of-sight $\nhsym$) or a more standard torus-like geometry. Alternatively, it would require a different soft X-ray contributor than is normally invoked for a more local source.

 Other possible contributors to this emission are photoionized gas, which is typically observed in local AGN \citep{BGC2006,GB2007}, or galaxy-scale collisionally ionized halos, which are usually observed in local ultraluminous galaxies \citep{TV2010,J2012,V2014,F2015,T2015}. Based on $f_{sc}$ for local low-luminosity AGN, photoionized gas can account for the soft excess. However, this would require a photoionized gas phase emitting with a luminosity on the order of $10^{44}$~\lumcgs\ , which is a factor of five to one order of magnitude higher than any scattered or photoionized gas phase ever measured in local luminous quasars in the X-rays \citep[][]{P2008,T2013,LM2016}. The recent stacking analysis carried out on high-z  EQR by \citet{G2018} reported a similarly high ($10^{44}$~\lumcgs) scattered component. They used the same {\sc MYTor} model as we did, with same assumptions on $\Gamma$ and torus inclination angle. In order to understand whether  these luminous scattered components are indeed associated with photoionized gas, a spectrum with a much higher signal-to-noise ratio is needed, hence it is not possible to make such a strong claim.
However, there are indications that the photoionized [OIII]-emitting gas that is usually associated with the X-ray emitting gas tends to be lacking or weak in a large fraction (40-70\%) of luminous AGN \citep[][]{N2004,SH2014,V2018} and is generally more compact than suggested from low-luminosity AGN \citep{N2004,BP2017}. This has been interpreted as some kind of departure from the phenomenology reported for less luminous sources \citep{N2004} and may indicate that the photoionized gas origin is unlikely.

We also evaluated the possibility that the soft-excess emission is due to a collisionally ionized plasma halo. The presence of the ionized Fe line, if confirmed at higher significance than what has been found in P15, may be indicative of high-temperature plasma. If we use a thermal model ({\sc APEC} in XSPEC) to parameterize the soft excess, we indeed estimate a plasma temperature of $kT\approx4$~keV. Together with the high luminosity of the soft excess, this may indicate intracluster-medium emission. This result may be tantalizing as hot DOGs have been claimed to be located in overdense galaxy regions \citep{J2014}. However, several caveats prevent us from making such a bold claim. The temperature may be driven by the presence of the low-significance putative ionized Fe line reported in P15, whose origin cannot be determined with the current data. Furthermore, as a first-order approximation, the cutoff of the thermal emission and the presence of L-shell Fe lines (a strong temperature indicator for $< 2$~keV plasma) cannot be determined for energies lower than $\sim$2~keV (as we cannot probe rest-frame energies much lower than this value). This means that our data are insensitive to lower temperatures, and we are not able to fully evaluate the parameter space down to lower temperatures.

\subsection{Nature of W1835}
Recently, \citet{farrah2017} performed a morphological analysis on {\it HST/WFC3} images of a sample of $z=2-3$  hot DOGs including W1835 in the optical rest-frame band (B band). They found W1835 to have an irregular light profile with no clear close companions. They employed several morphological indicators to quantify its appearance. They found that compared to all the other systems in their sample, the source is less likely to be in a clear merger state according to the boundaries in light concentration, asymmetry, and variance of the brightest 20\% of the galaxy light \citep{C2003,L2004,L2008}. It is therefore possible that either the system is in a non-merger state or in an advanced merger state, showing a common envelope surrounding an unresolved sub-kiloparsec scale nuclear region. \citet{farrah2017} concluded that there is no need to invoke a preferential link of hot DOGs with mergers, rather ascribing their phenomenology to brief and luminous nuclear accretion episodes in the evolutionary history of a massive star-forming $z\sim2$ galaxy. A similar analysis carried on by \citet{F2016b} instead concluded on a different sample (with very little overlap and including W1835) that hot DOGs are likely a transitional obscured phase in the merger-driven evolutionary QSO formation sequence, leading to the unobscured quasar phase. 
Interestingly, \citet{R2017b} found from analyzing the X-ray spectra of a sample of luminous and ultra-luminous infrared galaxies in different merging stages that CT AGN that are preferentially hosted in late-stage mergers, with the maximum obscuration reached when the galaxies have projected distances of 0.4-10.8~kpc. In the context of a merger-induced obscuration scenario, W1835 would therefore be interpreted as late-stage merger in which the nucleus is being obscured by infalling matter as a consequence of the loss of angular momentum of the ISM due to the merger phenomenon. However, unless there is a non-AGN-related prominent soft-excess contributor in our X-ray spectrum (i.e., photoionized or collisionally ionized copious amounts of gas), the obscuration cannot be ascribed to an homogeneous $4\pi$ obscurer that completely enshrouds the nucleus, as implied by quasar-induced merger formation scenarios, because the scattered fraction for the spherical obscurer is too high compared to what is typically observed in AGN.

\section{Conclusions}
We presented a $\sim155$~ks \nustar\ observation of the hot DOG W1835. The source was detected with a significance of $3.3\sigma$. 
We extracted \nustar\ spectra and jointly modeled them with the \xmm\ spectra that were previously presented in P15. We used both phenomenological and physically motivated models. The latter includes an appropriate treatment of the Compton scattering and accounts for the geometry of the obscurer. We explored two scenarios: 1) an edge-on torus, and 2) a sphere isotropically covering the nucleus. We find that\begin{itemize}
\item in all cases, a nearly CT ($\lognhsq\sim24$) to heavy CT ($\lognhsq \gg 24$) obscuration is required. This makes W1835 the first and most obscured $z>2$ AGN detected by \nustar\ so far; 
\item W1835 is very luminous and is a Compton-thick quasar, as indicated by the derived unobscured X-ray luminosity $\lumh\approx1-3\times10^{45}$~\lumcgs ($\lumhh\approx0.5-2\times10^{45}$~\lumcgs);  
\item a soft excess at low energies ($<2$~keV), which may amount to 5-15\% of the emission of the primary continuum is also necessary. The uniform spherical model is disfavored because the soft excesses it produces is too high (i.e., $>10\%$), unless a patchy obscurer distribution and/or an uncommon and powerful ($\sim10^{44}$~\lumcgs) non-AGN-related component is invoked.
\end{itemize}

We further investigated W1835 by performing optical-to-FIR SED modeling with a clumpy two-phase dusty toroidal model accounting for the MIR reprocessed AGN primary emission with the addition of a modified blackbody to model the FIR dust emission primarily heated by stellar processes. In order to be as consistent as possible with the X-ray analysis, we employed similar toroidal and spherical geometrical configurations for the MIR emitter. We found that
\begin{itemize}
\item the source is hyperluminous with a bolometric luminosity of $\lbol\approx3-5\times10^{47}$~\lumcgs;
\item the bolometric correction at 2-10~keV (10-40~keV) is on the order of  $\sim300$ ($\sim$500) and is consistent with those estimated in hyperluminous type~I AGN;   
\item it exhibits a powerful starburst with derived SFRs on the order of $\sim$3000~$M_\odot yr^{-1}$;
\item the ratio of X-rays to MIR is higher than the average value for the hot DOGs and is consistent within the uncertainties and accounting for all the modelings with those inferred in type~I hyperluminous AGN at the same cosmic epoch.
\end{itemize}

Considering the heavy obscuration, the luminosity, the ratio of X-rays to MIR, the SFR, and its \textit{HST} mildly disturbed morphology, this hot DOG can be interpreted as a late-stage merger in the context of the merger-induced quasar formation scenario. It exhibits relatively high values of soft excess, which, if confirmed, may indicate large reservoirs of photoionized or collisionally ionized gas. 

Future deep imaging observations of W1835 at higher spatial/spectral resolution both in X-ray (i.e., \chandra\ and {\it ATHENA}) and at submillimeter wavelengths ({\it NOEMA}) may enable us to further shed light on the nature of W1835. 

\begin{acknowledgements}
We thank the anonymous referee for the useful comments and suggestions. 
  LZ, EP, CV, and SB acknowledge financial support under ASI/INAF contract I/037/12/0.
RV acknowledges funding from the European Research Council under the European Union's Seventh Framework Programme (FP/2007-2013) / ERC Grant Agreement n. 306476. 
RM acknowledges ERC Advanced Grant 695671 ``QUENCH'' and support by the Science and Technology Facilities Council (STFC).  CR acknowledges financial support from FONDECYT 1141218, Basal-CATA PFB--06/2007, the China-CONICYT fund.

    This research has made use of the NASA/ IPAC Infrared Science Archive, which is operated by the Jet Propulsion Laboratory, California Institute of Technology, under contract with the National Aeronautics and Space Administration.
\end{acknowledgements}

%
%

\bibliographystyle{aa}
\bibliography{hotdog_}

\end{document}